\documentstyle[12pt,epsf]{article}
\def\Z0{${\em Z^0\/}$}

\def\r#1 {$^{#1}$}

\newcommand{\et}{{\rm E}_{\scriptscriptstyle\rm T}}

\newcommand{\ppbar}{p\bar{p}}

\newcommand{\ttbar}{t\bar{t}}
\newcommand{\bbbar}{b\bar{b}}

\def\gepsfcentered#1{
  \def\testit{#1}
  \def\lbracket{[}
  \ifx\testit\lbracket
    \let\dofilecmd=\gepsfwithopt
  \else
    \let\dofilecmd=\gepsfnoopt
  \fi
  \dofilecmd}

\def\gepsfnoopt#1{
  \begin{center}
  \leavevmode
  \epsffile{#1}
  \end{center}}

\def\gepsfwithopt#1 #2 #3 #4]#5{
  \begin{center}
  \leavevmode
  \gepsfmaxx=0.94\textwidth
  \epsffile[#1 #2 #3 #4]{#5}
  \end{center}}

\newdimen\gepsfmaxx
\def\epsfsize#1#2{
  \ifnum \epsfxsize=0
    \ifnum \epsfysize=0
      \ifnum #1 > \gepsfmaxx
        \gepsfmaxx
      \else
        #1
      \fi
    \else
      \epsfxsize
    \fi
  \else
    \epsfxsize
  \fi
}

\def \mgev   {GeV/$c^{2}$}
\def \pgev   {GeV/$c$}

\def \hhht     {$H_{T}$}
\def \et     {$E_{T}$}

\def \pt     {$P_{T}$}

\def \miset  {$\,/\!\!\!\!E_{T}$}

\def \ttbar  {$t\bar{t}$}
\def \ppbar  {$p\bar{p}$}

\def \bbbar  {$b\bar{b}$}

\renewcommand{\thebibliography}[1]{\subsubsection*{References}\list
  {[\arabic{enumi}]}{\settowidth\labelwidth{[#1]}\leftmargin\labelwidth
    \advance\leftmargin\labelsep\usecounter{enumi}}}

\begin{document}

\begin{flushright}
\begin{small}
hep-ex/9711004\\
\today\\
\end{small}
\end{flushright}

\vskip 0.2 in

\begin{center}
{\large\bf 
Observation of Hadronic $W$ Decays\\
 in \ttbar\ Events with the Collider Detector at Fermilab }
\vspace{2.0cm}

Submitted to PRL	
\end{center}

\font\eightit=cmti8
\def\r#1{\ignorespaces $^{#1}$}
\hfilneg
\begin{sloppypar}
\noindent
F.~Abe,\r {17} H.~Akimoto,\r {39}
A.~Akopian,\r {31} M.~G.~Albrow,\r 7 A.~Amadon,\r 5 S.~R.~Amendolia,\r {27} 
D.~Amidei,\r {20} J.~Antos,\r {33} S.~Aota,\r {37}
G.~Apollinari,\r {31}
% ...
\end{sloppypar}
\vskip .026in
\begin{center}
(CDF Collaboration)
\end{center}

\vskip .026in
\begin{center}
\r 1  {\eightit Argonne National Laboratory, Argonne, Illinois 60439} \\
\r 2  {\eightit Istituto Nazionale di Fisica Nucleare, University of Bologna,
I-40127 Bologna, Italy} \\
\r 3  {\eightit Brandeis University, Waltham, Massachusetts 02254} \\
\r 4  {\eightit University of California at Los Angeles, Los 
Angeles, California  90024} \\  
\r 5  {\eightit University of Chicago, Chicago, Illinois 60637} \\
\r 6  {\eightit Duke University, Durham, North Carolina  27708} \\
\r 7  {\eightit Fermi National Accelerator Laboratory, Batavia, Illinois 
60510} \\
\r 8  {\eightit University of Florida, Gainesville, FL  32611} \\
\r 9  {\eightit Laboratori Nazionali di Frascati, Istituto Nazionale di Fisica
               Nucleare, I-00044 Frascati, Italy} \\
\r {10} {\eightit University of Geneva, CH-1211 Geneva 4, Switzerland} \\
\r {11} {\eightit Harvard University, Cambridge, Massachusetts 02138} \\
\r {12} {\eightit Hiroshima University, Higashi-Hiroshima 724, Japan} \\
\r {13} {\eightit University of Illinois, Urbana, Illinois 61801} \\
\r {14} {\eightit Institute of Particle Physics, McGill University, Montreal 
H3A 2T8, and University of Toronto,\\ Toronto M5S 1A7, Canada} \\
\r {15} {\eightit The Johns Hopkins University, Baltimore, Maryland 21218} \\
\r {16} {\eightit Institut f\"{u}r Experimentelle Kernphysik, 
Universit\"{a}t Karlsruhe, 76128 Karlsruhe, Germany} \\
\r {17} {\eightit National Laboratory for High Energy Physics (KEK), Tsukuba, 
Ibaraki 305, Japan} \\
\r {18} {\eightit Ernest Orlando Lawrence Berkeley National Laboratory, 
Berkeley, California 94720} \\
\r {19} {\eightit Massachusetts Institute of Technology, Cambridge,
Massachusetts  02139} \\   
\r {20} {\eightit University of Michigan, Ann Arbor, Michigan 48109} \\
\r {21} {\eightit Michigan State University, East Lansing, Michigan  48824} \\
\r {22} {\eightit University of New Mexico, Albuquerque, New Mexico 87131} \\
\r {23} {\eightit The Ohio State University, Columbus, OH 43210} \\
\r {24} {\eightit Osaka City University, Osaka 588, Japan} \\
\r {25} {\eightit Universita di Padova, Istituto Nazionale di Fisica 
          Nucleare, Sezione di Padova, I-36132 Padova, Italy} \\
\r {26} {\eightit University of Pennsylvania, Philadelphia, 
        Pennsylvania 19104} \\   
\r {27} {\eightit Istituto Nazionale di Fisica Nucleare, University and Scuola
               Normale Superiore of Pisa, I-56100 Pisa, Italy} \\
\r {28} {\eightit University of Pittsburgh, Pittsburgh, Pennsylvania 15260} \\
\r {29} {\eightit Purdue University, West Lafayette, Indiana 47907} \\
\r {30} {\eightit University of Rochester, Rochester, New York 14627} \\
\r {31} {\eightit Rockefeller University, New York, New York 10021} \\
\r {32} {\eightit Rutgers University, Piscataway, New Jersey 08855} \\
\r {33} {\eightit Academia Sinica, Taipei, Taiwan 11530, Republic of China} \\
\r {34} {\eightit Texas A\&M University, College Station, Texas 77843} \\
\r {35} {\eightit Texas Tech University, Lubbock, Texas 79409} \\
\r {36} {\eightit Istituto Nazionale di Fisica Nucleare, University of Trieste/
Udine, Italy} \\
\r {37} {\eightit University of Tsukuba, Tsukuba, Ibaraki 315, Japan} \\
\r {38} {\eightit Tufts University, Medford, Massachusetts 02155} \\
\r {39} {\eightit Waseda University, Tokyo 169, Japan} \\
\r {40} {\eightit University of Wisconsin, Madison, Wisconsin 53706} \\
\r {41} {\eightit Yale University, New Haven, Connecticut 06520} \\
\end{center}

\vspace{2.0cm}

\begin{abstract}
	We observe 
hadronic $W$ decays in \ttbar\ $\rightarrow$
$W (\rightarrow \ell\nu)$ + $\ge$ 4 jet 
events using a 109 pb$^{-1}$ data sample of \ppbar\ collisions at $\sqrt{s}$ =
1.8 TeV collected with the Collider Detector at Fermilab (CDF). 
A peak in the dijet invariant mass distribution is obtained that is 
consistent with 
 $W$ decay and inconsistent with the background prediction by 
3.3$\sigma$. From this peak we measure the $W$ mass to be
77.2 $\pm$ 4.6 (stat+syst) \mgev. This result demonstrates the presence of
two $W$ bosons in \ttbar\ candidates in the 
$W (\rightarrow \ell\nu) + \ge$ 4 jet channel.


\end {abstract}

\newpage
\hspace{\parindent}

        CDF presented the first direct evidence for top quark production
 with 19.3 pb$^{-1}$ of data collected
in 1992-93~\cite{prd_uff} and also reported a detailed kinematic
study of top decays with this data sample~\cite{prd}.
In 1994-95,
the existence of the top quark was firmly established by
CDF~\cite{NEWPRL}
and D\O~\cite{D0}. CDF observed and studied \ttbar\ production
using a sample of 67~pb$^{-1}$ employing techniques similar to
those previously published~\cite{kinem,hanalysis}.

In the Standard Model, a top quark decays predominantly into a bottom quark and a $W$ boson.
As a part of the CDF kinematic studies of \ttbar\
production and decay, we have searched for
hadronic $W$ decays in a \ttbar\,-enriched sample containing a high-\pt\ lepton
 and four or more jets, which is consistent with the process 
$ p\bar{p} \rightarrow t\bar{t} X \rightarrow W^{+}bW^{-}\bar{b} X
 \rightarrow \ell \nu b jj \bar{b} X  $
with $\ell$ = $e$ or $\mu$, and $j$ representing a jet. Hereafter we call these
events ``$W+\ge 4$ jet events''.
The observation of hadronic $W$ decays in
$W+\ge 4$ jet events demonstrates the presence of two $W$ bosons 
in the final state for \ttbar\ candidates.

	The studies reported here
 use a $109 \pm 7~{\rm pb}^{-1}$ data sample.  Three analysis techniques yield
 signal-to-background ratios ranging from 0.23 to 1.4 in the dijet 
invariant mass region
 60-100~\mgev.

	The CDF detector consists of a magnetic spectrometer surrounded by calorimeters
and muon chambers~\cite{detector}. A silicon vertex detector (SVX), located 
immediately outside the beampipe, provides precise track reconstruction
in the plane transverse to the beam and is used to identify secondary vertices
from $b$ and $c$ quark decays~\cite{svx}. The momenta of charged particles 
are measured in the central tracking chamber (CTC) which is 
in a 1.4~T superconducting solenoidal magnet. Outside the CTC, electromagnetic
and hadronic calorimeters cover the pseudorapidity~\cite{cdf_def} region
$|\eta|<$ 4.2 and are used to identify jets and electron candidates.
Outside the calorimeters, drift chambers in the region $|\eta|<$ 1.0 provide
muon identification.

	To select \ttbar\ $\rightarrow$ $W+\ge 4$ jet candidates, we require
an isolated, high-\pt\ electron or muon (\pt\ $>$ 20 \pgev), high missing
transverse energy~\cite{cdf_def} (\miset\ $>$ 20 GeV),
 three jets with \et\ $>$ 15 GeV and $|\eta|<$ 2.0,
 and a fourth jet with \et\ $>$ 8 GeV and $|\eta|<$ 2.4.
After these cuts 163 events remain.
The expected fraction of \ttbar\ events in this sample is 33\%.
The fraction was determined by extrapolating  the background calculation performed on
the  $W+\ge 3$ jet sample for the \ttbar\ cross section measurement~\cite{cross_section}
to include the additional requirement
of a fourth jet.

For the event selection, neither the jet \et\ nor the \miset\ are corrected
for detector effects.
However, when calculating a dijet invariant mass, 
jet energies are corrected by a pseudorapidity and energy-dependent 
factor which accounts for such effects as calorimeter nonlinearity, reduced
response at detector boundaries, contributions from the underlying event
and multiple interactions and 
losses outside the clustering cone~\cite{jetcor}.
We apply an additional
energy correction to the four highest-\et\ jets in a \ttbar\ candidate event.
The correction depends on the type of parton they are assigned to: 
a light quark, a hadronically decaying $b$ quark, or a $b$ quark that
decayed semileptonically~\cite{prd_uff}.
This parton-specific
correction was derived from a study of \ttbar\ events
generated with the HERWIG Monte Carlo program~\cite{herwig}.
This correction is applied in order to reconstruct the original parton 
energy as closely as possible.
These cuts and corrections are the same as those used in the top mass 
measurement~\cite{prd_uff} before requiring a $b$ jet.

	Monte Carlo $t\bar{t}$ events are generated by 
the HERWIG program~\cite{herwig}
with a top mass of 175 \mgev~\cite{new_mass}. 
	The  expected 
background from direct $W +$ jets production
is estimated using the VECBOS Monte Carlo calculation~\cite{vecbos} 
to generate $W +$ 3 parton matrix elements.
To model $W+\ge 4$ jet events we implement a simulation of parton fragmentation
with a shower algorithm based on HERWIG.
The lowest-order matrix elements are sensitive to the choice of the
mass scale in the strong coupling constant $\alpha$$_{s}$.
We use two $Q^{2}$ scales, namely,
 the square of the average jet \pt\ 
($\langle P_{T}\rangle ^{2}$) and the square of the $W$ boson mass 
($M_{W}^{2}$). We use $\langle P_{T}\rangle ^{2}$  as a standard $Q^{2}$
scale and $M_{W}^{2}$ for an estimate of the systematic uncertainty.
A more detailed description of the Monte Carlo samples
can be found elsewhere~\cite{prd_uff}.

\begin{figure}[htbp]
 \vspace{2cm}
 \epsfysize=15cm
  \epsffile[0 0 200 600]{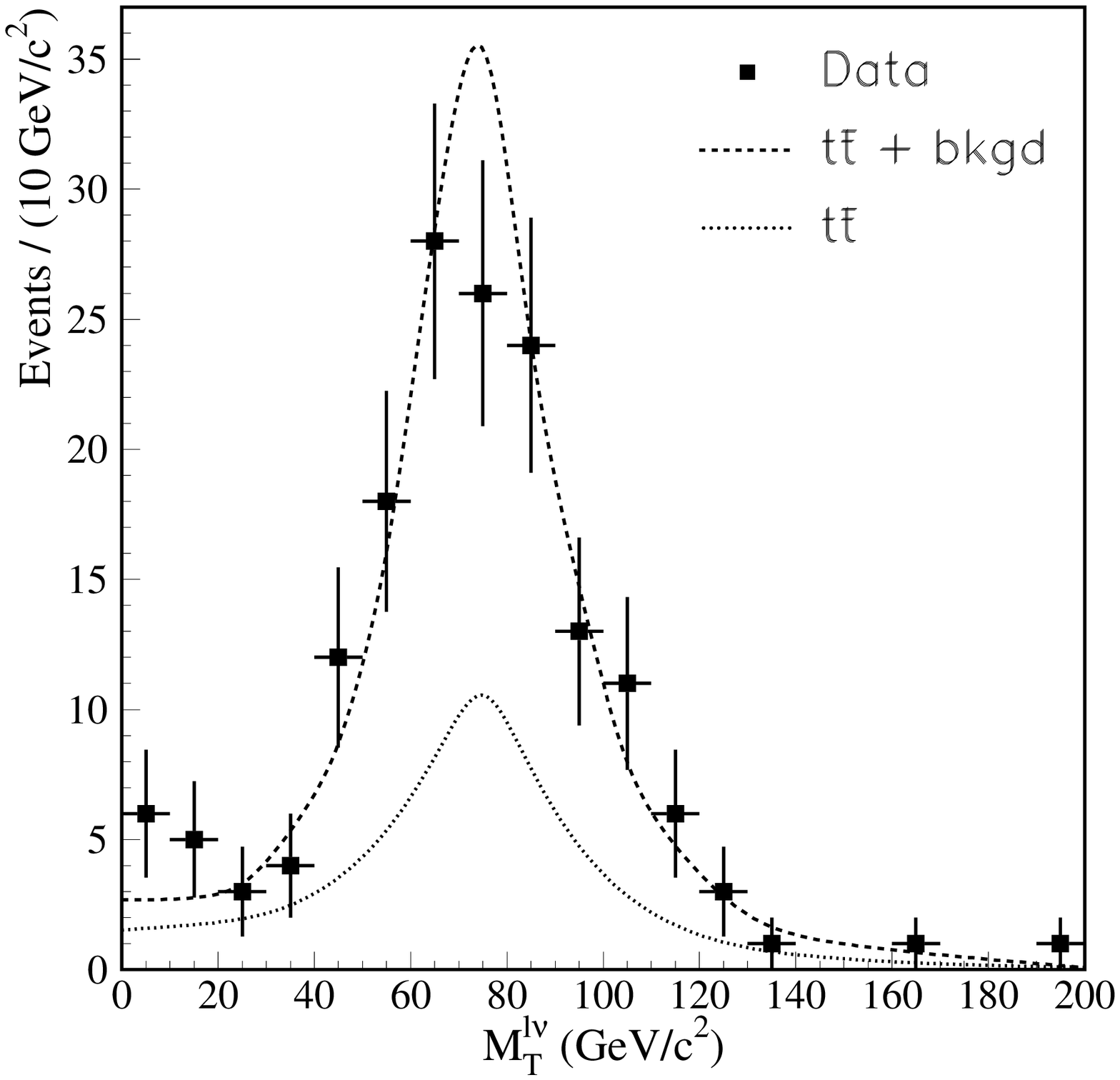}
 \vspace{-3cm} \hspace{2cm}
\caption{ The lepton-\miset(neutrino) transverse mass distribution
in the $W+\ge 4$ jet data (points)
along with the Monte Carlo expectations from \ttbar\ (dotted) 
and \ttbar\ plus QCD $W$ + jets background (dashed).
}
\label{fig0}
\end{figure}

The lepton-\miset\ transverse mass distribution is shown
in Figure~\ref{fig0}.  Also shown are the Monte Carlo 
expectations from \ttbar\ as well as
\ttbar\ plus QCD $W$ + jets background.
The Monte Carlo expectations are normalized to the number of events
observed.

We use three methods to search for $W$ decay to two jets:
a top mass reconstruction technique,
a total transverse energy cut, 
and the identification of both $b$ jets. 

	In the first method,
we search for a $W$ decaying to two jets by kinematically reconstructing 
\ttbar\ $\rightarrow \ell \nu b jj \bar{b}$ with a constrained
fitting technique similar to the one used for the determination of the top 
mass~\cite{prd_uff}.
The constraints used in the top mass measurement
include the requirement that the two jets hypothesized to come from the $W$
decay have an invariant mass equal to the $W$ mass.  
Here we eliminate the $W$ mass constraint.
There are multiple solutions due to the ambiguity in 
determining the neutrino longitudinal momentum and the assignment of jets
to the parent partons.
We choose the solution with the lowest $\chi^2$.
The resolution and signal-to-background ratio 
for \ttbar\ events is improved by requiring a $b$-tag
in the event and requiring that the tagged jet correspond
to a $b$ jet in the fit.  
To tag $b$ jets, we exploit either the long lifetime of $b$ quarks
by requiring a secondary vertex (SVX $b$-tag),
or the semileptonic
decays of $b$ quarks
by searching for additional leptons (SLT $b$-tag)~\cite{prd_uff}.
The fraction of correct $W$ jet assignments was found from Monte Carlo studies to be improved 
from 22\% before $b$-tagging
to 37\% after tagging.  

\begin{figure}[htbp]
 \vspace{2cm}
 \epsfysize=15cm
  \epsffile[0 0 200 600]{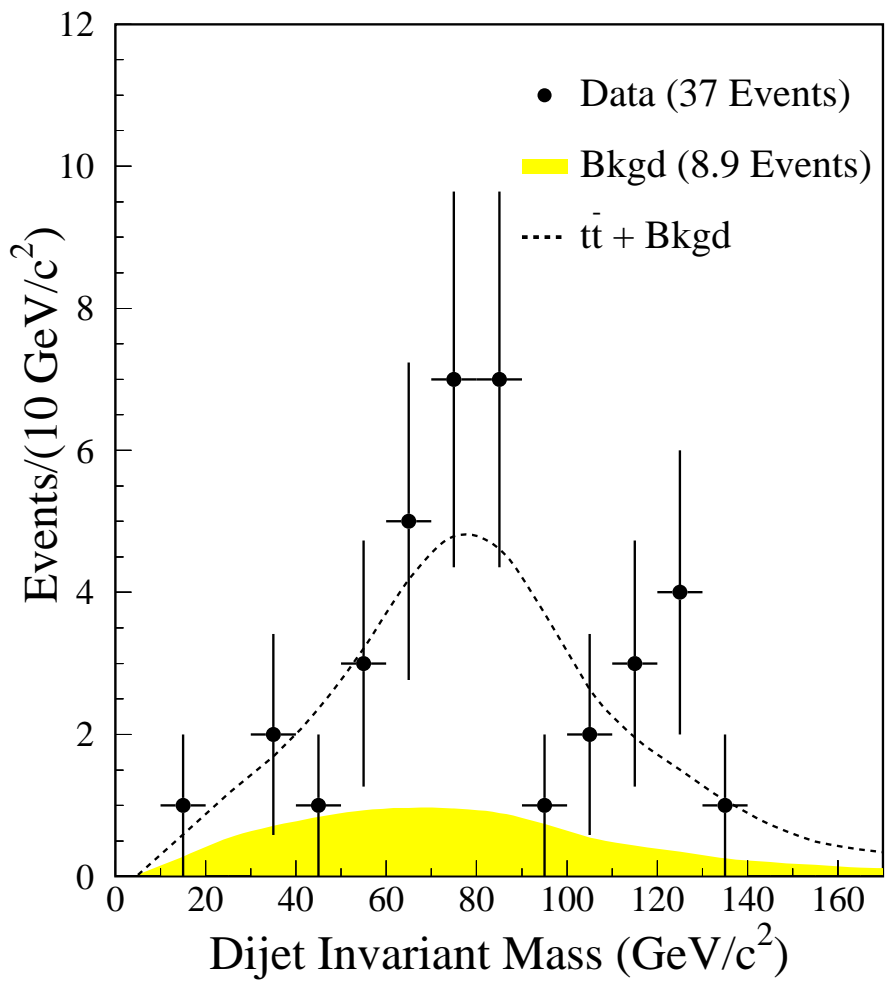}
 \vspace{-3cm} \hspace{2cm}
\caption{
Invariant mass distribution 
for the two jets that are associated with the $W$ by
the constrained kinematic fit. The shaded histogram
shows the expected distribution from background
events; \ttbar\ plus background events are shown as a dashed line.}
\label{fig3}
\end{figure}

	Of the 163 events passing the 
$W + \geq$ 4 jet selection cuts, 
37 have at least one $b$-tag found by the SVX or 
SLT algorithms.  The expected background is $8.9^{+2.0}_{-1.7}$ events, calculated using 
techniques described in~\cite{cross_section}.
 The dijet invariant mass distribution 
for these events is shown in Figure~\ref{fig3}.
Also shown are the Monte Carlo distributions for $t\bar{t}$ and background,  
normalized to
28.1 and 8.9 events respectively.  We note a strong enhancement in
the data distribution near the world-average $W$ mass~\cite{wmass}.


	The second method employs a cut on the minimum
total transverse energy \hhht\ to enhance the $t\bar{t}$ contribution
in the sample.
\hhht\ is defined as the scalar sum of the \pt\ of the lepton (\et\ for an electron),
the four highest-\et\ jets, and the \miset\ \cite{hanalysis}. 
We optimize the \hhht\ threshold (310 GeV) using \ttbar\ and background
simulations. This cut improves the fraction of \ttbar\ events in the sample
from 33\% to 55\%.

We calculate the invariant mass of each two-jet pair out of the four
highest-\et\ jets.  Monte Carlo simulations predict that two jets from $W$ decay
are included in the four highest-\et\ jets with an efficiency of 66\%.
There are six combinations of two jets in each event.
To determine the contributions from non-$t\bar{t}$ backgrounds and combinatoric
backgrounds in $t\bar{t}$ events, we fit the dijet mass
distribution $M_{jj}$ outside the $W$ mass region 
to the following function:
$N_{bg} f_{bg}(M_{jj}) + N_{comb} f_{comb}(M_{jj})$,
where $N_{bg}$ and $N_{comb}$ are the number of non-\ttbar\ 
and \ttbar\ combinatoric
background dijet pairs respectively, 
and $f_{bg}$ and $f_{comb}$ are their dijet mass distributions, derived
from Monte Carlo.
Since non-\ttbar\ backgrounds, such as $WW$, $WZ$, and \bbbar\,
are smaller and give dijet mass distributions similar to QCD $W$ + jets,  
$f_{bg}$ is obtained from VECBOS calculation of QCD $W$ + jets background.

\begin{figure}[htbp]
 \vspace{2cm}
 \epsfysize=15cm
  \epsffile[0 0 200 600]{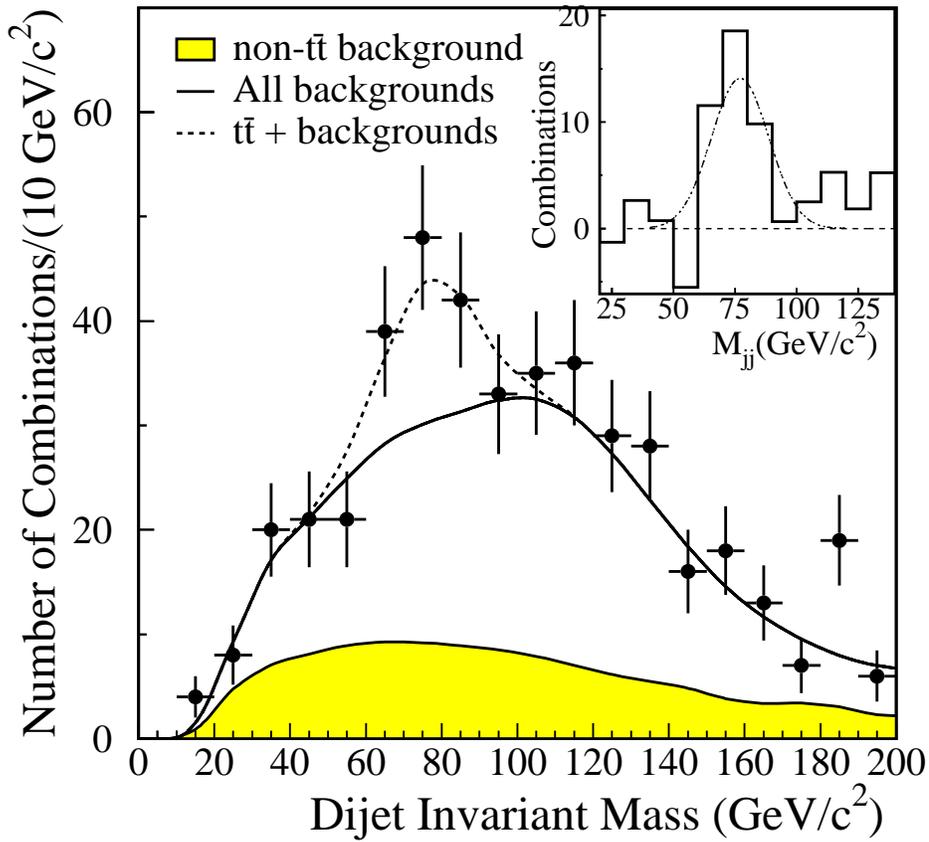}
 \vspace{-3cm} \hspace{2cm}
\caption{
Dijet mass distribution in $ W+\geq$ 4 jet events
after the $H_{T}$ $>$ 310 GeV cut.
Also shown are a fitted curve of the sum (solid)
of the VECBOS $W$ + jets background (shaded) and
the combinatoric background from \ttbar\ events.
A Gaussian distribution 
fitted to the $W$ mass peak is shown by the dotted curve.
The inset shows the $W$ mass peak after subtracting the backgrounds.}
\label{fig2}
\end{figure}

We extract 
the $W \rightarrow$ 2 jets signal by 
subtracting the QCD $W+$ jets background and 
the combinatoric background as shown in Figure~\ref{fig2}.
For the  $W+\ge 4$ jet sample of \ttbar\ production,
the reconstructed mass
peak of the $W$ decaying into dijets is well described by
a Gaussian distribution with a standard deviation of 11.7 \mgev.
 We fit this distribution to a Gaussian function with a fixed width
and obtain $29 \pm 13$ $W \rightarrow $ 2 jet events in the mass region 70-90 \mgev,
which is consistent with the
number of $ W \rightarrow $ 2 jet events ($20 \pm 6$) 
expected from the CDF \ttbar\ production cross section
measurement~\cite{cross_section}.
This excess is inconsistent with the expected background by 2.8$\sigma$,
corresponding to a probability of 2.6 $\times$ 10$^{-3}$ of it being
a background fluctuation. 
The signal-to-background ratio in the mass region 60-100 \mgev\ is 0.23, and the overall
fraction of non-\ttbar\ background is ($27 \pm 14$)\%.
The dijet invariant mass peak has a mean of 
77.1 $\pm$ 3.8(stat) $\pm$ 3.6(syst) \mgev. 
The systematic uncertainty comes mainly from the jet energy scale,
which is the uncertainty on how well our measured jet energies correspond
to the original parton energies.  The systematics are
listed in Table~\ref{tab:uncer3}, 
and are described in more detail in Ref.~\cite{new_mass}. 


        In the third method, we isolate the hadronic $W$ decay by
identifying two $b$ jets in the
four highest-\et\ jets of $W + \geq 4$ jet events. Such events are hereafter
called ``double $b$-tagged'' $W + \geq 4$ jet events.
Double $b$-tagging eliminates ambiguity about which two jets
are assigned to the $W$,
and further suppresses non-\ttbar\ background. 
  Once we find at least one $b$ jet with an SVX tag or an SLT tag
in an event, we look for the second $b$ jet. The second $b$-tag can be
an SVX tag, an SLT tag or a Jet Probability tag~\cite{aleph}.
The Jet Probability algorithm obtains the probability
that a jet is consistent with the decay of a zero-lifetime particle
by using the impact parameters of the tracks in the jet as measured
in the silicon vertex detector.
Jets with probabilities less than 5\% are considered tagged.
We form the invariant mass of the remaining two untagged
jets.

In our HERWIG $t\bar{t}$ Monte Carlo sample, 25\% 
of the events are double $b$-tagged.
The technique finds the two correct $W$ jets in 43\% of these
events.  However, this number is sensitive to the amount of initial
and final state radiation in the simulation, since the largest 
contamination (38\% of double $b$-tagged events)
comes from events where the
four highest-\et\ jets do not correspond to
two $b$ jets and two jets from $W$ decay.  
Other $t\bar{t}$ backgrounds include events where
both $W$ bosons decayed leptonically and only one lepton was identified (8\%),
events where a $c$ quark from $W$ decay was tagged (8\%), and
mistags in jets from $W$ decay (3\%), where a mistag means a tag of a jet
originating from neither a $b$ nor a $c$ quark.

We expect 1.3~$\pm$~0.3 double-tagged events from non-$t\bar{t}$ background,
with the largest contributions being 0.6~$\pm$~0.2 events
from mistags and
0.4~$\pm$~0.2 events from $Wb\bar{b}$ and $Wc\bar{c}$ processes which 
are QCD $W +$ jets events containing real heavy flavor.

\begin{figure}[htbp]
 \vspace{2cm}
 \epsfysize=15cm
  \epsffile[0 0 200 600]{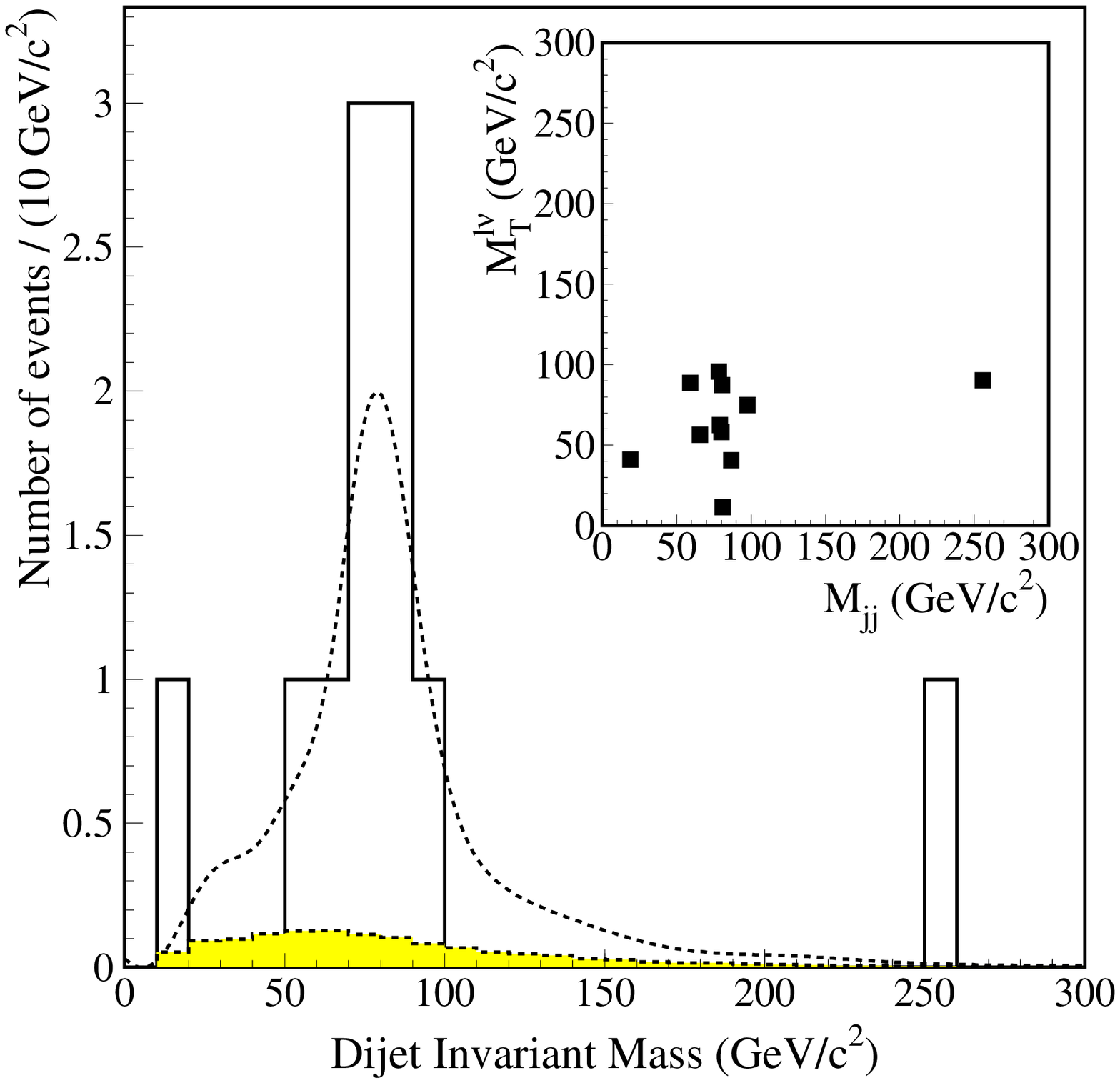}
 \vspace{-3cm} 
\hspace{2cm}
\caption{Dijet mass spectrum of the 
two untagged jets in double $b$-tagged events.
The shaded curve shows the expected distribution from QCD $W$ + jets background 
and the dashed curve is from \ttbar\ plus background.
The inset compares
the lepton-\miset(neutrino) transverse mass to the dijet invariant mass
in these eleven events showing evidence for events with both 
a leptonically-decaying
$W$ and a hadronically-decaying $W$.}
\label{mjj_2b_data}
\end{figure}

In the data, we find eleven 
double $b$-tagged events.
The dijet invariant mass of the two untagged
jets in these eleven events is shown in Figure~\ref{mjj_2b_data}.
Eight of the eleven dijet
combinations fall in the mass window of 60-100 \mgev.
In Monte Carlo simulations of both QCD $W$ + jets and $t\bar{t}$ backgrounds,
only about a third of the dijet
mass combinations fall in this window. 
The inset plot in Figure~\ref{mjj_2b_data} shows the mass of the hadronic $W$
candidates against the transverse mass of the leptonic $W$ candidates in the
same events.

We fit this mass distribution to a sum of a Gaussian
$W \rightarrow$ 2 jets signal, $t\bar{t}$ backgrounds, 
and non-$t\bar{t}$ backgrounds, and  
obtain a $W$ mass of 78.1 $\pm$ 4.4(stat) $\pm$ 2.9(syst) 
\mgev, 
with the systematic uncertainties listed in Table~\ref{tab:uncer3}.
The fitted $W$ signal has 8.7 events, and
0.7 events from $t\bar{t}$ backgrounds.
Constraining the fit to the expected  $W \rightarrow$ 2 jets 
fraction in $t\bar{t}$ events
yields 6.4 $W$ events, and a $W$ mass of 78.3 $\pm$ 5.1(stat) $\pm$ 
2.9(syst) GeV/c$^2$. We use this value for obtaining a final combined 
$W$ mass.

We use likelihood
fits to evaluate the significance of this peak.  
We first fit the data using a Gaussian $W$ signal 
whose mean is the world-average $W$ mass plus \ttbar\ and non-\ttbar\
backgrounds.
Next we remove the signal term
and note the change in likelihood.  We then repeat this
procedure on Monte Carlo pseudo-experiments
using only \ttbar\ and non-\ttbar\ background events, 
and study how often the likelihood
changes by at least the amount seen in the data.  
This change in likelihood was seen with a probability of 1.7 $\times$ 10$^{-3}$,
corresponding to a Gaussian significance of 2.9$\sigma$.
If we construct the pseudo-experiments with the
expected $W$ signal fraction,
we find that a change in likelihood larger than the one seen in the
data occurs 15\% of the time.


\begin{table}[htbp]
\begin{center}
\begin{tabular}{lcccc}
\hline\hline
Uncertainty (\mgev) & \hhht\ cut & double $b$-tag & common & combined\\ \hline
Statistical	& 4.6	& 5.1	& - &3.5 \\
\\
Jet energy scale & 	& 	& &2.8\\
(a) Detector effects   &	1.9  & 1.9 & 1.9 &\\
(b) Soft gluon effects  &  2.0 & 2.0 & 2.0 &\\
\\
Other systematics & &  &&0.6\\
(a) Backgrounds   &  0.5 & 0.2 & - & \\
(b) Fitting		&  1.9	& 0.6 & - & \\
(c) \hhht\ cut		&  1.0  & - & - & \\
\\
Total uncertainty & 5.8 & 5.9 & 2.8 & 4.6\\ \hline\hline
\end{tabular}
\end{center}
 \caption{\label{tab:uncer3} The list of uncertainties (in  \mgev\ )
 in the $W \to$ 2 jets mass for the \hhht\ cut and double $b$-tag analyses.
 The columns ``common'' and ``combined'' correspond to the uncertainties common in 
 the two analyses
 and the combined uncertainties of the two analyses, respectively.}
\end{table}
 
An overall result is obtained by excluding the eleven
double $b$-tagged events from the \hhht\ cut analysis.
We then combine the two analyses to obtain a $W$ mass peak significance of 
3.3$\sigma$, corresponding to a
probability of 5.4 $\times$ 10$^{-4}$ of it being a background fluctuation
and a $W$ mass of 77.2 $\pm$ 3.5(stat) $\pm$ 2.9(syst) \mgev\ with
the systematic uncertainties listed in Table~\ref{tab:uncer3}.
This is consistent with the current world-average $W$ mass of 80.375 $\pm$
0.120 \mgev~\cite{wmass}.

	In conclusion, we observe
hadronic $W$ decays as a dijet mass peak in 
 \ttbar\ $\rightarrow$ $W + \ge$ 4 jet events.
This demonstrates the presence of two $W$ bosons in the \ttbar\
	candidates.
In the future, these techniques can be used to set limits on nonstandard top decays.

	We thank the Fermilab staff and the technical staffs of the 
participating institutions for their contributions.  This work was supported
by the U.S. Department of Energy and the National Science Foundation;
the Italian Istituto Nazionale di Fisica Nucleare; 
the Ministry of Science, Culture, and Education of Japan; 
the Natural Sciences and Engineering Research Council of Canada; 
the National Science Council of the Republic of China and
the A. P. Sloan Foundation.

\newpage

\newpage

\newpage


\begin{thebibliography}{99}
\bibitem{prd_uff} 
F. Abe $et~al$., 
  Phys. Rev. Lett. {\bf 73}, 225 (1994), hep-ex/9405005;
F. Abe $et~al$.,  
  Phys. Rev. {\bf D 50}, 2966 (1994).
\bibitem{prd}
F. Abe $et~al$.,
  Phys. Rev. {\bf D 51}, 4623 (1995), hep-ex/9412009.
\bibitem{NEWPRL} 
F. Abe $et~al$., 
Phys. Rev. Lett. {\bf 74}, 2626 (1995), hep-ex/9503002.
\bibitem{D0} 
S. Abachi $et~al$.,  Phys. Rev. Lett. {\bf 74}, 2632 (1995),  hep-ex/9503003.
\bibitem{kinem} 
F. Abe $et~al$.,
  Phys. Rev. {\bf D 52}, R2605 (1995).
\bibitem{hanalysis} 
F. Abe $et~al$., Phys. Rev. Lett. {\bf 75}, 3997 (1995), hep-ex/9506006.
\bibitem{detector}
F. Abe $et~al$.,
Nucl. Instrum. Methods Phys. Res., Sect. {\bf A 271}, 387 (1988).
\bibitem{svx}
S. Cihangir  $et~al$., Nucl. Instrum. Methods Phys. Res., Sect. {\bf A 360}, 137 (1995).
Our previous silicon vertex detector is described in D. Amidei $et~al$.,
Nucl. Instrum. Methods Phys. Res., Sect. {\bf A 350}, 73 (1994).
\bibitem{cdf_def} In the CDF coordinate system, $\theta$ is the polar angle 
with respect to the proton beam direction. The pseudorapidity $\eta$ is defined 
as $-$ln tan($\theta$/2). The transverse momentum of a particle is \pt\ = 
$P\sin\theta$. The analogous quantity using
calorimeter energies, \et\ = $E\sin\theta$, is
called transverse energy.
Missing
transverse energy \miset\ is defined as $- \Sigma E_{T}^{i}\cdot \hat{n}_{i} $,
where $\hat{n}_{i}$ are the unit vectors, in the plane transverse to the
beam line, pointing from the interaction point to the energy deposition
in cell $i$ of the calorimeter.
\bibitem{cross_section} F. Abe $et~al$., FERMILAB-PUB-97/286-E, hep-ex/9710008.
\bibitem{jetcor} F. Abe $et~al$., Phys. Rev. {\bf D 45}, 1448 (1992);
F. Abe $et~al$., Phys. Rev. {\bf D 47}, 4857 (1993).
\bibitem{herwig} G. Marchesini and B. R. Webber, Nucl. Phys. {\bf B 310}, 
461 (1988);
G. Marchesini $et~al$., Comput. Phys. Comm. {\bf 67}, 465 (1992).
\bibitem{new_mass}
F. Abe $et~al$., FERMILAB-PUB-97/284-E.
\bibitem{vecbos} F. A. Berends, W. T. Giele, H. Kuif, B. Tausk, Nucl. Phys.
{\bf B 357}, 32 (1991);\\
W. Giele, Ph.\,D.~Thesis, Leiden (1989).
\bibitem{wmass} 
M. Lancaster, Proceedings of de la Vall\'ee d'Aoste, La Thuile, 1997; 
ed. M. Greco.
\bibitem{aleph} D. Buskulic $et~al$., Phys. Lett. {\bf B 313}, 535 (1993).
\end{thebibliography}
\end{document}